\documentclass[11pt]{article} 
\usepackage{authblk} 

\usepackage[utf8]{inputenc} 

\usepackage{geometry} 
\usepackage{graphicx} 

\usepackage{booktabs} 
\usepackage{array} 
\usepackage{paralist} 
\usepackage{verbatim} 
\usepackage{subfig} 

\usepackage{mathtools}

\usepackage{tikz}

\usepackage{tikz-feynman}

\usepackage{fancyhdr} 
\usepackage{sectsty}
\allsectionsfont{\sffamily\mdseries\upshape} 

\usepackage[nottoc,notlof,notlot]{tocbibind} 
\usepackage[titles,subfigure]{tocloft} 

\begin{document}

\title{Meaningful Details: The value of adding baseline dependence to the Neutrino-Dark Matter Effect}
\author{William S. Marks and Fu-Guang Cao}
 \affil{Institute of Fundamental Sciences, Massey University, \\
 Private Bag 11 222, Palmerston North, New Zealand}

\date{} 

\maketitle
\begin{abstract}
The possible effect on the flavour spectra of astronomical neutrinos from a neutrino-dark matter interaction has been investigated for decoherent neutrinos \cite{DeSalas2016}.
In this work, we report results calculated for coherent neutrinos.  This was done with two different models for the neutrino dark-matter interactions: a flavour state interaction, 
as for the weak interaction in the Standard Model, and a mass state interaction, which is predicted by certain non-Standard Models (specifically Scotogenic models). 
It was found that using a coherent analysis dramatically increased the explorable parameter space for the neutrino-dark matter interaction.  
However, the detection of coherent astronomical neutrinos presents a significant challenge to experimentalists, because such a detection would require an improvement in energy resolution by at least six orders of magnitude, with similar improvements in astronomical distance determinations.
\end{abstract}
\section{Introduction}

Astronomical observations over the last several decades have consistently shown that, in simple terms, galaxies spin too fast for the visible matter they contain to hold them together gravitationally.  This means that either the standard theory of gravity needs to be modified or that there are other forms of matter present which do not participate in the other Standard Model interactions (the electroweak and strong interactions) or only do so extremely weakly.  Given that no satisfactory alternative theory of gravity has been developed, it is generally presumed that there is a type of matter that is quite common but is currently undetectable.  This matter is called ``Dark Matter" (DM) \cite{Lesgourgues2012,Duda2011}.  

Neutrinos are electro-magnetically neutral leptons with a very small mass (the heaviest neutrino is at least six orders of magnitude lighter than the electron) \cite{Mertens2016}.  Intrinsically connected to the massive nature of neutrinos is neutrino oscillations, also known as flavour mixing \cite{Bilenky2014}.  (Indeed, the massive nature of neutrinos was proven by the discovery of neutrino oscillations \cite{Ahmad2001,Collaboration2001}.)

The currently favoured model for neutrino oscillations is three-neutrino mixing whereby a neutrino produced in one of the three flavour states enters into a quantum superposition of the three mass states which then propogate at different velocities due to having different masses.  These velocity differences produce interference effects which cause the flavour content of the wavepacket to change in an oscillatory manner.

An interesting feature of these oscillations is that their exact form is dependent on the medium in which the oscillations take place.  The most common form is the Mikheyev-Smirnov-Wolfenstein (MSW) effect, which occurs when neutrinos pass through a medium containing electrons \cite{Wolf1978,Barger,Renshaw2013}.  The coupling of the electron-neutrino component to the medium causes a shift in the oscillation pattern, which is observable if the density of the medium is high enough.  Theoretically, a similar effect would appear for any medium that couples to neutrinos strongly enough, regardless of the precise interaction involved.

De Salas, Lineros and, T{\'o}rtola \cite{DeSalas2016} proposed that this effect could be used to observe neutrino-dark matter interactions.  The idea was that if there is a significantly strong coupling between neutrinos and Dark Matter, then it would be observable at neutrino observatories, particularly high energy ones like IceCube \cite{Aartsen2013}.  One attractive feature of their approach is that it does not presume any particular type of Dark Matter or any particular type of interactions, merely that there is Dark Matter and it couples to neutrinos.

The formula for flavour transition probabilities when the wavepackets are coherent is:
\begin{equation}
f_{\beta}=\displaystyle\sum_{\alpha=e,\mu,\tau}{\left|\displaystyle\sum_{i=1}^{3}U_{{\beta}i}^{}U_{i{\alpha}}^{\dagger}\mathrm{e}^{-\mathrm{i}E_{i}t}\right|}^2f_{\alpha},
\label{oscprob}
\end{equation}
where \(f_{\alpha}\) is the emission flavour vector (i.e. a normalized vector containing the three flavour fractions), \(f_{\beta}\) is the flavour vector at the detector
and U is the matrix translating between the flavour basis and the effective mass basis\footnote{A note about the effective mass basis:  The effective mass basis is simply the one in which the Hamiltonian in the presence of a medium is diagonal.  It is called the effective mass basis because, if it is calculated, then the equations for oscillations in a vacuum can be used with simple substitution.}.  De Salas, Lineros and, T{\'o}rtola \cite{DeSalas2016} used the formula for decoherent wavepackets:
\begin{equation}
f_{\beta}=\displaystyle\sum_{\alpha=e,\mu,\tau}\left(\displaystyle\sum_{i=1}^{3}{\left|U_{{\beta}i}^{}U_{{\alpha}i}^{*}\right|}^2f_{\alpha}^0\right),
\label{nooscprob}
\end{equation}
which can be obtained by expanding out the term in straight brackets in Eq.~\ref{oscprob} and then integrating over all \(t\) before recompressing the surviving terms.  Since baseline is related to time, this effectively integrates over all possible source locations.  Thus, the decoherent probability is an average of the coherent probability across all baselines.


The coherent formula is valid as long as the different mass state wavepackets overlap.  As the neutrinos propagate, the mass state wavepackets diverge due to minute differences in their group velocities \cite{Akhmedov2009}.  Eventually, the wavepackets diverge to the point where they no longer overlap.  When this happens, oscillations cease to occur as it is the interference between the wavepackets that produces oscillations; the neutrinos are now said to be decoherent and the valid formula is the decoherent one.  The time it takes for neutrinos to decohere increases with energy \cite{Akhmedov2012}.  Thus, neutrinos can maintain coherence, even across astronomical distances, if they have sufficiently high energy.  (Coherence limits are discussed in more detail in Section 2.)

This work examined the effects of using the coherent formula and how they differ from the decoherent formula.  This was done by performing calculations using both formulas in the same manner as in \cite{DeSalas2016}, save that the coherent formula required the addition of a baseline dependency.  The calculations were made using two different potentials, one corresponding to interactions with the neutrino flavour states (Section 3) and one corresponding to interactions with the neutrino mass states (Section 4).  The flavour based potential is the standard choice and corresponds to DM that participates in the weak interaction (e.g. WIMPs).  It was the potential used by \cite{DeSalas2016}.  The mass based potential is an unconventional choice that is predicted by certain theories of neutrino mass that are beyond the Standard Model (i.e. Scotogenic models) \cite{Wilkinson2014}.  These are models where neutrinos obtain mass radiatively via virtual intermediaries which can serve as DM candidates (see e.g. \cite{Ma2009,HiroshiOkada2014,Merle2015,Bonilla2016}).

In addition to comparing results from the coherent formula with those from the decoherent formula, the results for the two potentials were compared to each other using the coherent formula.  This is done by examining the distance dependency of the coherent formula for each potential (Section 5).  The analysis in Section 3  was then carried out for neutrinos from blazar TXS 0506+056 at the energy observed by IceCube \cite{Collaboration2018,2041-8205-854-2-L32} in order to examine the effect of an increased baseline (Section 6).  The implications of the analyses are discussed in Section 7 before the whole work is summarized in Section 8.

\section{Coherence conditions}

The coherent formula is valid as long as the different mass state wavepackets overlap, which occurs over a certain period of time.  Since time is related to distance travelled, this translates to a requirement that the baseline, \(L\), be much less than the coherence length, \(L_{coh}\) \cite{Akhmedov2009}, which is given by:
\begin{equation}
L \ll L_{coh} = \sigma_{X}\frac{V_G}{{\Delta}V} \approx \sigma_{X}\frac{2E^2}{\Delta{m}^2},
\label{lcohineq}
\end{equation}
where \(\sigma_{X}\) is the wavepacket size, \(V_G\) is the group velocity, \({\Delta}V\) is the difference in wavepacket velocities, and the ultra-relativistic approximations, \(V_G \approx 1\) and \({\Delta}V \approx \frac{\Delta{m}^2}{2E^2}\), have been used\footnote{\(\Delta{m}^2\) is the larger of the two mass-squared differences, as that is the dominant factor.  Also, the convention \(\hbar=c=1\) has been used.}.  An important feature to remember is that the wavepacket size is an effective size.  This is, in general, the larger of the source's physical size and the inverse of the detector's energy resolution (presuming that the two differ greatly).  Practically speaking, it is whichever makes observations the most difficult.

The limitation on the wavepacket size is that it be smaller than the oscillation wavelength, i.e:
\begin{equation}
\sigma_{X} \ll l_{osc} \approx \frac{4{\pi}p}{\Delta{m}^2},
\label{oscszlim}
\end{equation}
where \(p\) is the momentum of the wavepacket, which is generally substituted for energy, due to the ultra-relativistic nature of the neutrinos.  The wavepacket size will at least be as large as the source object.  This puts a limit on the maximum size of the source.  In this study, we presumed that intra-galactic high energy neutrinos are produced in the near vicinity of black holes, particularly stellar mass black holes.  Since these objects are fairly compact, this constraint is not considered an issue.\footnote{The black hole under consideration, A0620-00, is estimated to have a mass of \(6.6\pm0.25\) solar masses\cite{Cantrell2010}, which gives it a Schwarzschild radius of approximately 20 km.  By contrast, the short wavelength oscillations for the 1 PeV neutrinos under consideration are approximately 5-6 AU in length, with 1 AU \(\approx 1.5\times10^8\) km.  Even though it can be safely assumed that the production region is much larger than the Schwarzchild radius, it is still probably much smaller than the oscillation wavelength.}  Of course, from an experimental standpoint this constraint is significant in as much as the baseline needs to be determined to a precision that is smaller than the oscillation wavelength.

Equations \ref{lcohineq} and \ref{oscszlim} can be combined and rearranged to give
\begin{equation}
L \ll \frac{8{\pi}E^3}{(\Delta{m}^2)^2}
\end{equation}
which for \(E \geq 1 \mathrm{TeV}\) is satisfied by all baselines within the observable universe by several orders of magnitude.

In order to determine the relationship with the energy resolution, it helps to employ the relationship between wavepacket size and energy uncertainty,
\begin{equation}
\sigma_{X} \approx \frac{V}{\sigma_E} \approx \frac{1}{\sigma_E}.
\label{szenrel}
\end{equation}
Combining this relationship with inequalities, Eq.~\ref{lcohineq} and Eq.~\ref{oscszlim}, gives \cite{Akhmedov2012},
\begin{equation}
\frac{\Delta{m}^2}{4{\pi}E^2} \ll \frac{\sigma_E}{E} \ll \frac{2E}{L\Delta{m}^2}.
\label{erezlim}
\end{equation}
This puts a significant limit on the energy uncertainty for large \(L\).  At 1 PeV and with a baseline of 1 kpc (the parameters used in this work), the requirement on energy resolution is \(\frac{\sigma_E}{E} \ll 10^{-6}\).  We notice that this required energy resolution is much smaller than that available in current neutrino experiments.  For example, the energy resolution at IceCube is about \(10^{-1}\) and the upcoming JUNO neutrino observatory is slated to achieve an energy resolution of around \(10^{-2}\), albeit at a different energy range, with the resolution scaling inversely with the square root of energy \cite{JUNO2015}.\footnote{Since baseline will have an uncertainty, \(\sigma_L\), associated with it, the middle term in Eq.~\ref{erezlim} should be \(\frac{\sigma_E}{E}+\frac{\sigma_E}{E}\frac{\sigma_L}{L}\).  Given that the requirement on the baseline uncertainty was already established as \(\sigma_L \ll l_{osc}\), the correction of the energy resolution from baseline uncertainty is negligible.}

The main motivation of this work is to investigate what new physical insights can be obtained about the neutrino-DM interaction from use of the coherent formalism, in comparison with the results obtained using the decoherent formalism, in anticipation of currently unknown experimental technologies that enable such a small energy resolution to be achieved.  Furthermore, the possibly significant impacts on our understanding of both DM and neutrinos from such a study will provide strong motivations for experimentalists to overcome the challenges in achieving such an observation.
\section{Flavour Potential}
The Hamiltonian used for the flavour state interactions was:
\begin{equation}
\begin{matrix*}[l]
\mathcal{H}_{tot}&=\mathcal{H}_{vac}+U^{\dagger}\mathcal{H}_{int}U\\
&=
\begin{bmatrix}
E_1 & 0 & 0 \\
0 &  E_2 & 0 \\
0 & 0 &  E_3 \\
\end{bmatrix}+U^{\dagger}
\begin{bmatrix}
V_{11} & 0 & 0 \\
0 & V_{22} & 0 \\
0 & 0 & 0 \\
\end{bmatrix}U\mathrm{,}
\end{matrix*}
\end{equation}
where \(\mathcal{H}_{int}\) is taken to be Hermitian and \(U\) is the Pontecorvo-Maki-Nakagawa-Sakata (PMNS) matrix:
\begin{equation}
U=
\begin{bmatrix}
 c_{12}c_{13}& s_{12}c_{13} & s_{13}\mathrm{e}^{-\mathrm{i}\delta} \\
 -s_{12}c_{13}-c_{12}s_{23}s_{13}\mathrm{e}^{\mathrm{i}\delta}& c_{12}c_{23}-s_{12}s_{23}s_{13}\mathrm{e}^{\mathrm{i}\delta} & s_{23}c_{13} \\
 s_{12}s_{23}-c_{12}c_{23}s_{13}\mathrm{e}^{\mathrm{i}\delta}&-c_{12}s_{23}-s_{12}c_{23}s_{13}\mathrm{e}^{\mathrm{i}\delta} & c_{23}c_{13}
\end{bmatrix}
\times{diag(1,\mathrm{e}^{\mathrm{i}\frac{\alpha}{2}},\mathrm{e}^{\mathrm{i}\frac{\beta}{2}})}\mathrm{.}
\label{PMNS}
\end{equation}

Following the procedure laid out in \cite{Barger}, the flavour fractions (for the coherent formula) in the presence of a medium are calculated using equations analogous to the equations for vacuum oscillations.  In a vacuum (\(V_{11}=V_{22}=0\)), the flavour fraction at the detector, \(f_{\beta}\) is given by:
\begin{eqnarray}
f_{\beta}&=&f_{\alpha}\times((U_{\beta,1}^{}U^{\dagger}_{1,\alpha})^2+(U_{\beta,2}^{}U^{\dagger}_{2,\alpha})^2+(U_{\beta,3}^{}U^{\dagger}_{3,\alpha})^2 \nonumber \\
& &+2U_{\beta,1}^{}U^{\dagger}_{1,\alpha}U_{\beta,2}^{}U^{\dagger}_{2,\alpha}{\rm cos}(\Delta{E}_{12}t)+\dots \nonumber \\
& &+2U_{\beta,1}^{}U^{\dagger}_{1,\alpha}U_{\beta,3}^{}U^{\dagger}_{3,\alpha}{\rm cos}(\Delta{E}_{13}t)
+2U_{\beta,2}^{}U^{\dagger}_{2,\alpha}U_{\beta,3}^{}U^{\dagger}_{3,\alpha}{\rm cos}(\Delta{E}_{23}t))\mathrm{,}
\label{Eq:mixing_vacuum}
\end{eqnarray}
where \(\Delta{E}_{ij}\) is the difference between energies \(i\) and \(j\), \(t\) is the time since neutrino creation and \(U_{l,k}\) is the (\(l,k\)) term in the PMNS matrix, Eq.~\ref{PMNS}.  In the presence of a medium (some non-zero \(V_{ij}\)) and taking the ultra-relativistic limit (\(E_{i}\approx\frac{m^2_{i}}{2E}\) and \(t\) is replaced with \(L\), the baseline), this is modified to:
\begin{eqnarray}
f_{\beta}&=&f_{\alpha}\times((W_{\beta,1}^{}W^{\dagger}_{1,\alpha})^2+(W_{\beta,2}^{}W^{\dagger}_{2,\alpha})^2+(W_{\beta,3}^{}W^{\dagger}_{3,\alpha})^2 \nonumber \\
& &+2W_{\beta,1}^{}W^{\dagger}_{1,\alpha}W_{\beta,2}^{}W^{\dagger}_{2,\alpha}{\rm cos}(\frac{\Delta{M}^2_{12}}{2E}L)+\dots \nonumber \\
& &+2W_{\beta,1}^{}W^{\dagger}_{1,\alpha}W_{\beta,3}^{}W^{\dagger}_{3,\alpha}{\rm cos}(\frac{\Delta{M}^2_{13}}{2E}L)
+2W_{\beta,2}^{}W^{\dagger}_{2,\alpha}W_{\beta,3}^{}W^{\dagger}_{3,\alpha}{\rm cos}(\frac{\Delta{M}^2_{23}}{2E}L))\mathrm{,}
\label{Eq:mixing_DM}
\end{eqnarray}
where \(\Delta{M}^2_{ij}\) is the effective mass-squared difference (which is energy dependent) and \(W_{l,k}\) is the (\(l\),\(k\)) term in the effective mixing matrix, which is also energy dependent.  These terms are derived from the diagonalization of \(\mathcal{H}_{tot}\) to produce \(\mathcal{H}_{eff}=W^{\dagger}\mathcal{H}_{tot}W\), the effective Hamiltonian, which was done numerically.\footnote{Since the diagonalization was done numerically, the actual computations replaced \(\frac{\Delta{M}^2_{13}}{2E}\) with \(\delta\lambda_{ij}\), the difference between diagonal terms \(\lambda_{i}\) and \(\lambda_{j}\) in \(\mathcal{H}_{eff}\).  Thus, Eq.~\ref{oscprob} becomes 
\begin{displaymath}
f_{\beta}=\displaystyle\sum_{\alpha=e,\mu,\tau}{\left|\displaystyle\sum_{i=1}^{3}W_{{\beta}i}^{}W_{i{\alpha}}^{\dagger}\mathrm{e}^{-\mathrm{i}\lambda_{i}t}\right|}^2f_{\alpha}.
\end{displaymath}} 
It should be clear that decoherence sets the cosine terms in Eq.~\ref{Eq:mixing_DM} to zero, leaving only the first three terms in the equation.

In \cite{DeSalas2016}, all parameters in \(\mathcal{H}_{int}\) were set to zero except for \(V_{11}\) and \(V_{22}\).  Setting \(V_{33}\) to zero was justified by renormalization.  The values ranged from \(\pm10^{-23}\) eV to \(\pm10^{-3}\) eV.  The physicality of the negative potentials is justified by the fact that only differences between the parameters matter in the calculations and it is always possible to subtract a scalar multiple of the identity matrix from the Hamiltonian, as the final result will remain the same.  

This study used the same general form as \cite{DeSalas2016} (only \(V_{11}\) and \(V_{22}\) are non-zero).  However, the range was chosen to be  \(10^{-33}\) eV to \(10^{-13}\) eV in order to better coincide with the potentials allowed by cosmological considerations \cite{Mangano2006,Serra2009,Wilkinson2014}.  Additionally, only the positive ranges were considered for reasons of convenience and as it suffices to describe the result (also, the results in \cite{DeSalas2016} were symmetric about the two axes).  As in \cite{DeSalas2016}, the starting flavour fraction composition was \((e,{\mu},{\tau})=(1,0,0)\).  The baseline was chosen to be \(1 \mathrm{kpc}\), which is approximately the distance to the nearest, suspected, black hole from Earth, A0620-00\cite{Cantrell2010}.  The energy was set to be 1 PeV as that was the highest energy examined by \cite{DeSalas2016}.

The metric used in \cite{DeSalas2016} to measure the effect was:
\begin{equation}
R_{\beta}=\frac{(f_{\beta}-f_{\beta,0})}{f_{\beta,0}},~\beta=e, \mu, \tau,
\label{Rdef}
\end{equation}
where \(f_{\beta}\) is the flavour fraction of flavour \(\beta\) in the presence of Dark Matter and \(f_{\beta,0}\) is the vacuum value.  The inclusion of the \(f_{\beta,0}\) term as a denominator at first brush appears to normalise the parameter, but a careful examination reveals that this is not the case.  First of all, it suppresses the deviation when \(f_{\beta}<f_{\beta,0}\) relative to when \(f_{\beta}>f_{\beta,0}\).  Secondly, it allows for the appearance of singularities.  A better parameter would be the simple difference:
\begin{equation}
D_{\beta}=f_{\beta}-f_{\beta,0},~\beta=e, \mu, \tau .
\end{equation}
This parameter was used in this work due to the appearance of near singular values in Eq.~\ref{Rdef} when the coherent function is used.

\begin{figure}
\includegraphics[width=\textwidth,trim={0 3.5cm 0 4.5cm},clip]{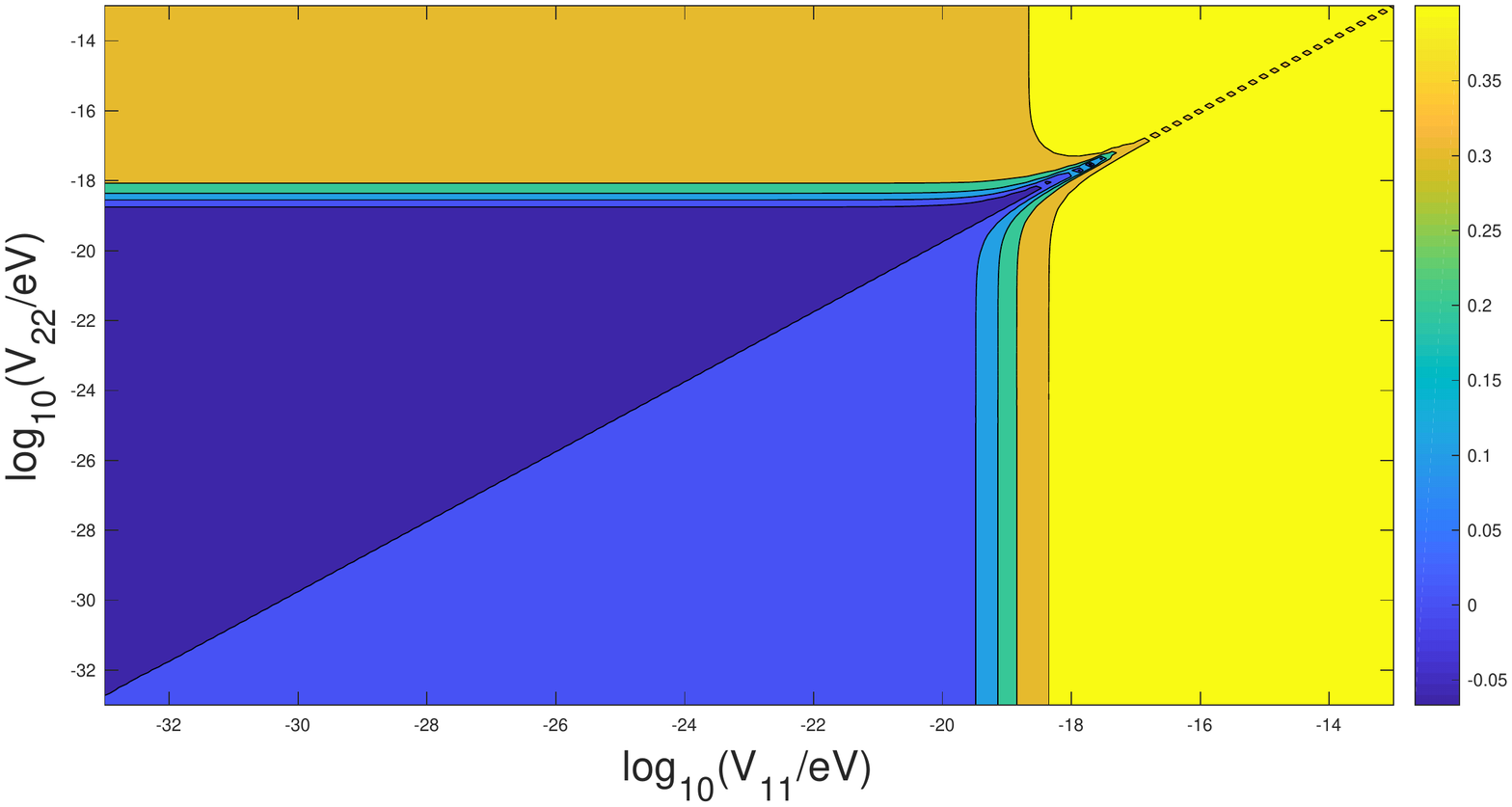}
\caption{This shows \(D_{e}\) across flavor potentials at \(1~\mathrm{PeV}\) using the decoherent formula, Eq.~\ref{nooscprob}.
The regions of substantial shift are clearly visible at the top and on the right-hand side while the region of little effect is in the lower left-hand corner. 
The difference in colour on either side of the null-line is due to a sign change across that line.  This is the same for all other plots.}
\label{SLTFBNORe}
\end{figure}

\begin{figure}
\includegraphics[width=\textwidth,trim={0 3.5cm 0 4.5cm},clip]{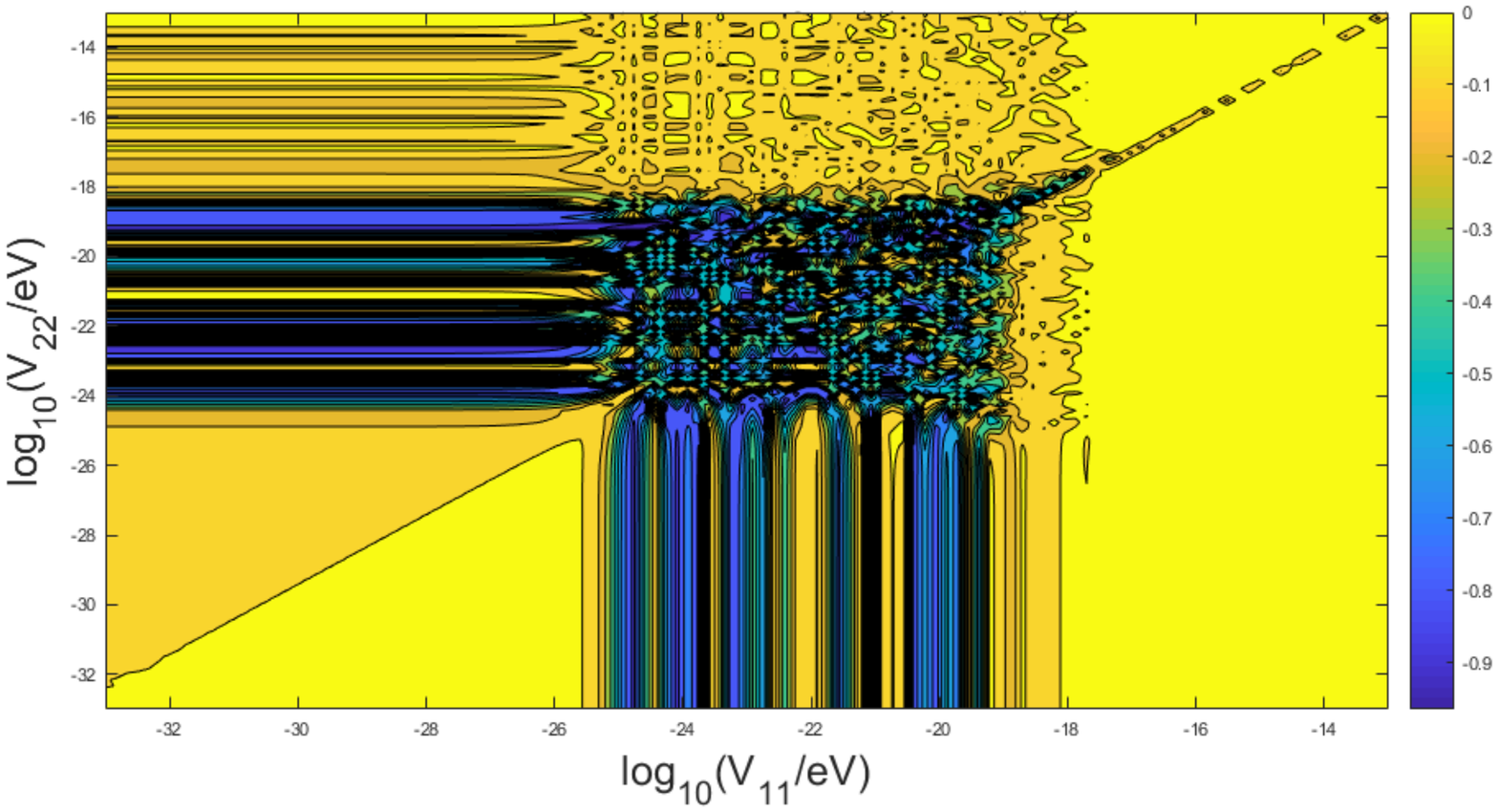}
\caption{This shows \(D_{e}\) across flavor potentials at \(1~\mathrm{PeV}\) using the coherent formula, Eq.~\ref{oscprob}. 
The baseline was \(L=3{\times}10^{19}~\mathrm{m} \approx 1~\mathrm{kpc}\).  Note that the color scale is different from the decoherent plot, as the value range is different and the software used to perform the calculations and generate the plots does not support manually defined colorscales on contour plots.  Also note that the high potential region has a different value from the same region in the decoherent plot.  That is because the value in that region for the coherent formula is still baseline dependent, with the decoherent result being the average of the coherent results for all baselines.}
\label{SLTFBWORe}
\end{figure}

\begin{figure}
\includegraphics[width=\textwidth,trim={0 3.5cm 0 4.5cm},clip]{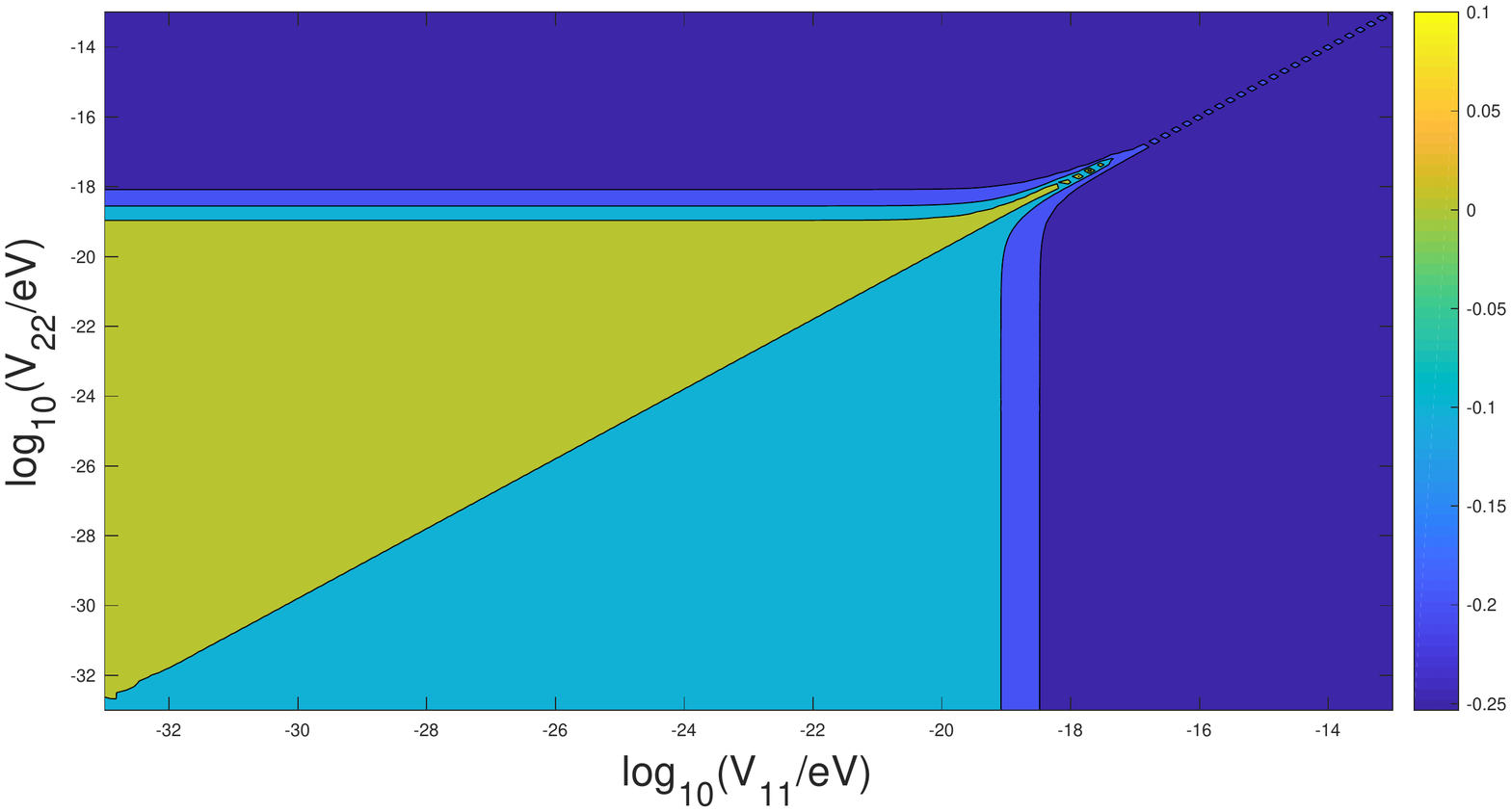}
\caption{This shows \(D_{\mu}\) across flavor potentials at \(1~\mathrm{PeV}\) using the decoherent formula, Eq.~\ref{nooscprob}.
The overall pattern is nearly identical to the pattern for electron neutrinos, which is expected due to the equivalence of the functions.}
\label{SLTFBNORmu}
\end{figure}

\begin{figure}
\includegraphics[width=\textwidth,trim={0 3.5cm 0 4.5cm},clip]{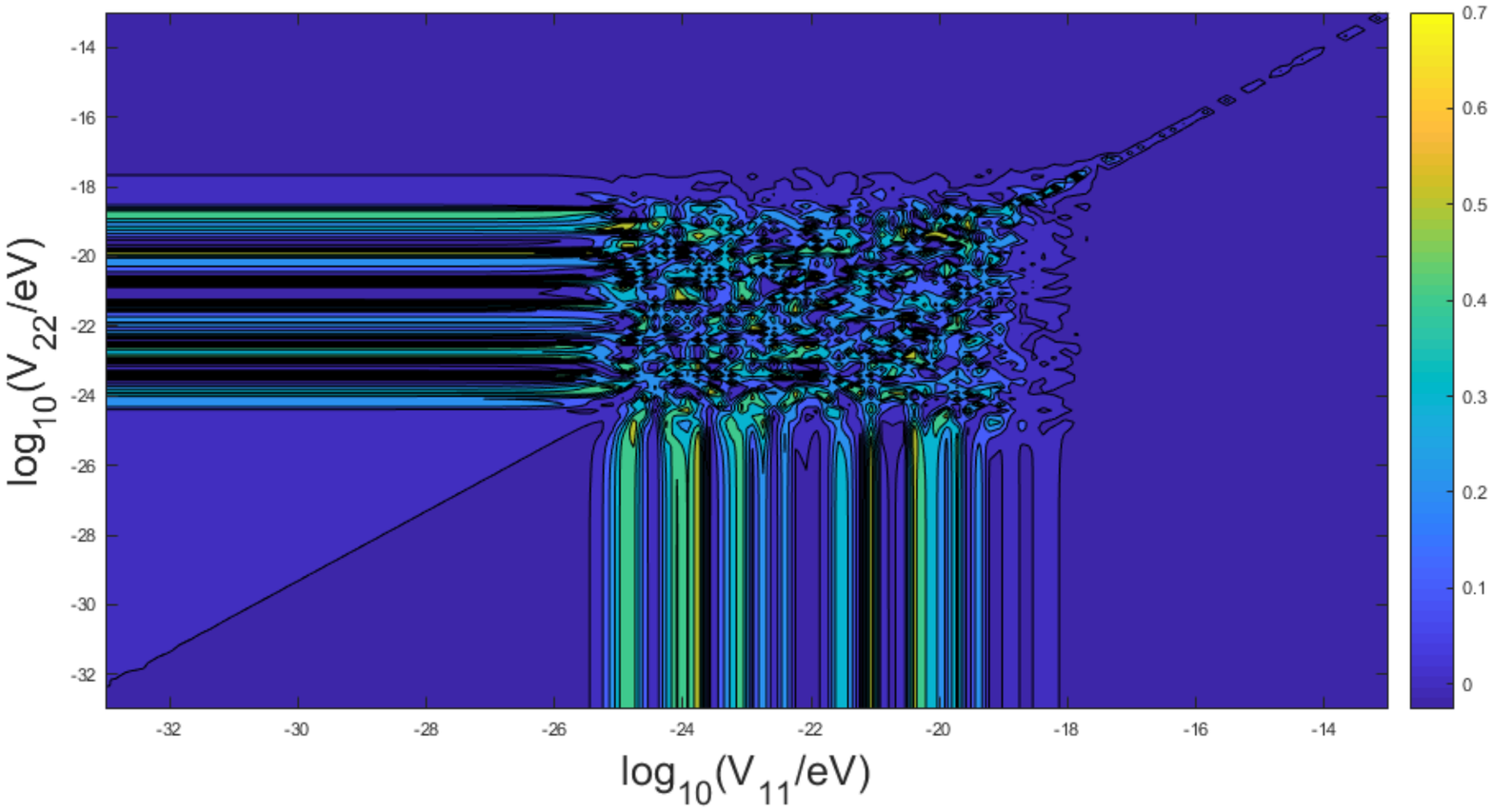}
\caption{This shows \(D_{\mu}\) across flavor potentials at \(1~\mathrm{PeV}\) using the coherent formula, Eq.~\ref{oscprob}. 
The baseline was \(L=3{\times}10^{19}~\mathrm{m} \approx 1~\mathrm{kpc}\).  Again, the pattern is largely the same as for the same case with electron neutrinos.}
\label{SLTFBWORmu}
\end{figure}

\begin{figure}
\includegraphics[width=\textwidth,trim={0 3.5cm 0 4.5cm},clip]{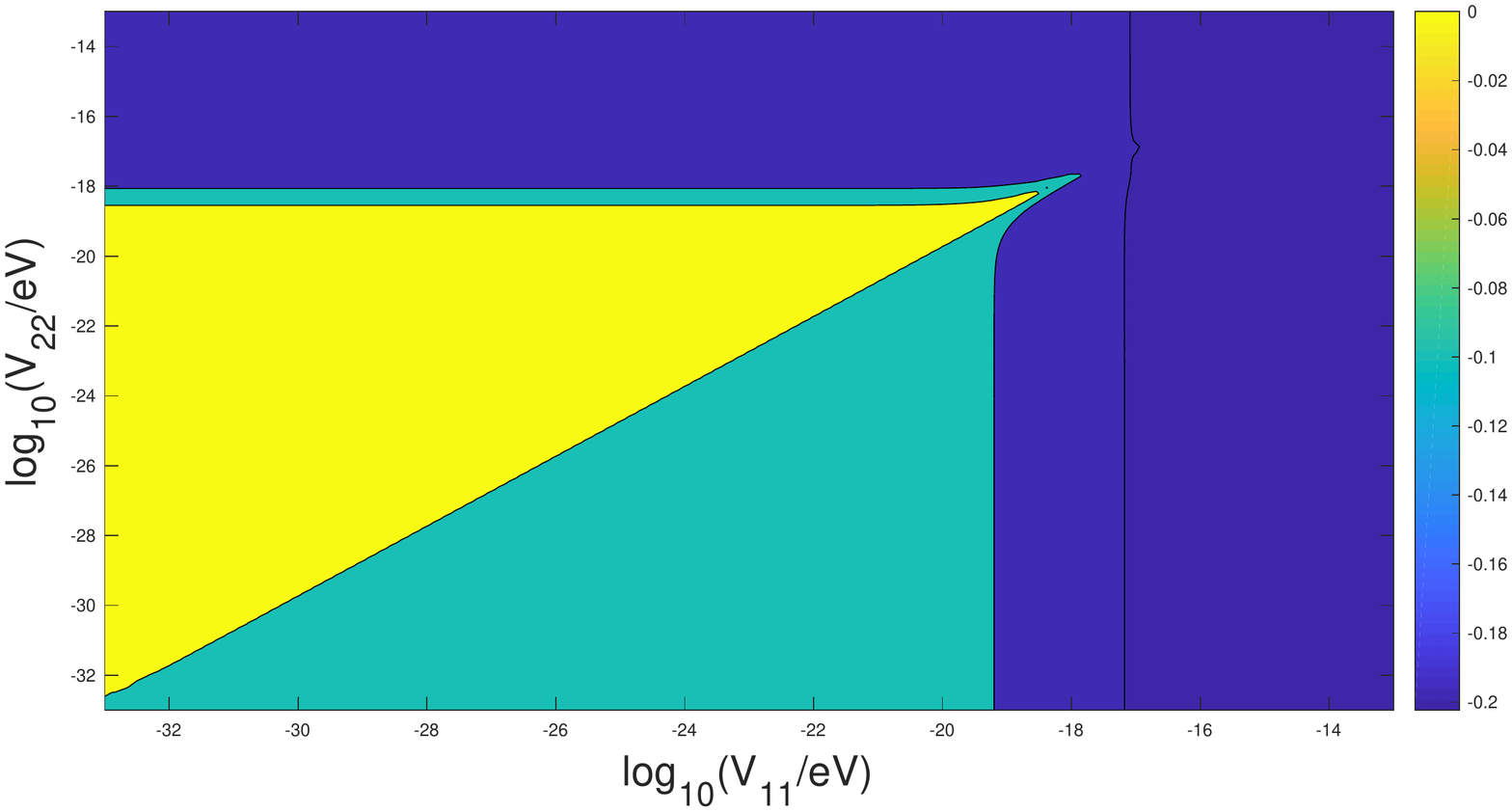}
\caption{This shows \(D_{\tau}\) across flavor potentials at \(1~\mathrm{PeV}\) using the decoherent formula, Eq.~\ref{nooscprob}.
 The behaviour is still largely the same as for the other two flavours.}
\label{SLTFBNORtau}
\end{figure}

\begin{figure}
\includegraphics[width=\textwidth,trim={0 3.5cm 0 4.5cm},clip]{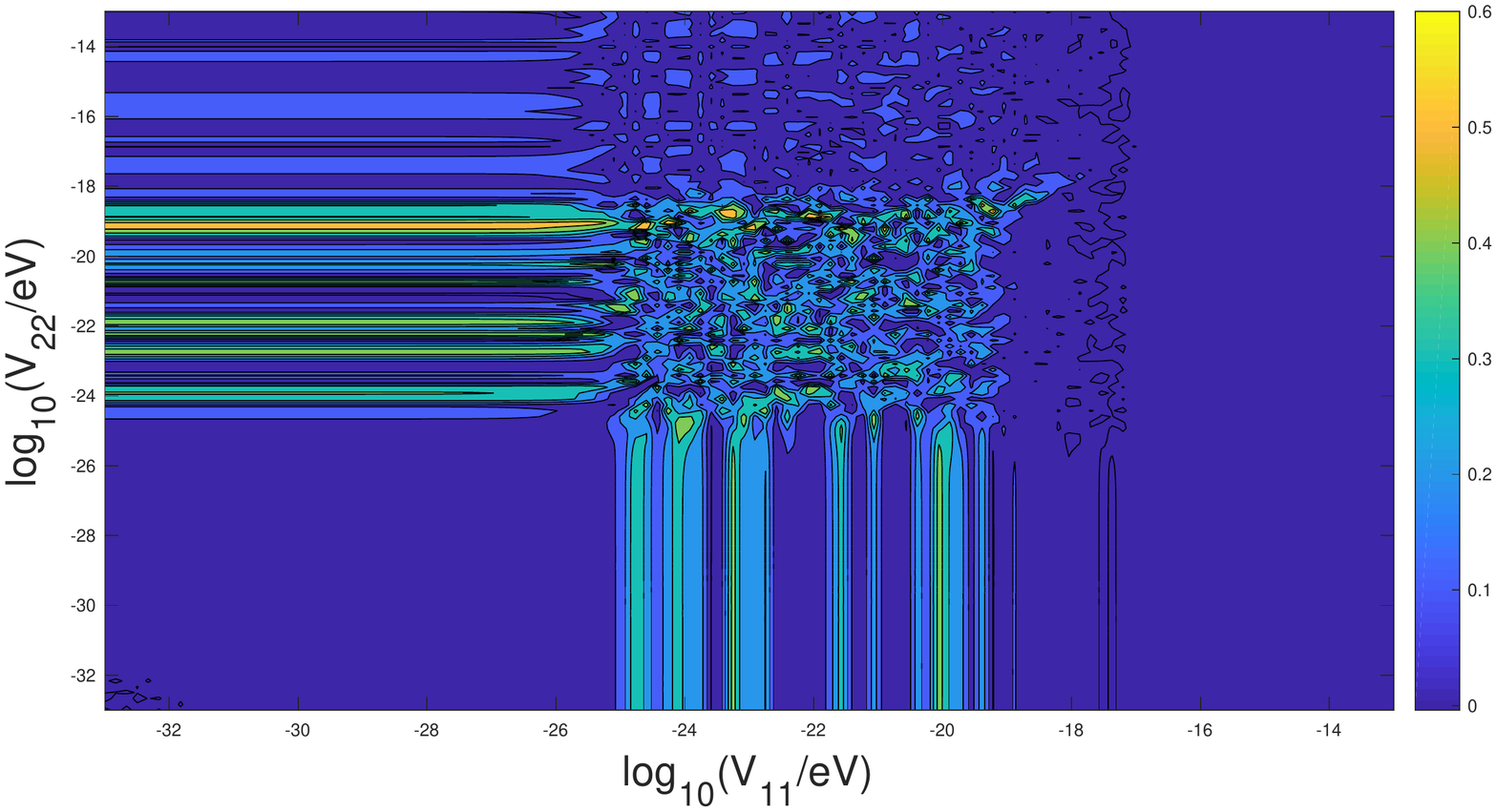}
\caption{This shows \(D_{\tau}\) across flavor potentials at \(1~\mathrm{PeV}\) using the coherent formula, Eq.~\ref{oscprob}.
The baseline was \(L=3{\times}10^{19}~\mathrm{m} \approx 1~\mathrm{kpc}\). 
Once again, the behaviour in the pattern is largely the same as for the other flavours using the same function, as expected.}
\label{SLTFBWORtau}
\end{figure}

The results for the decoherent function (Eq.~\ref{nooscprob}), shown in figs.~\ref{SLTFBNORe},\ref{SLTFBNORmu} and \ref{SLTFBNORtau}, for \(D_{e}\), \(D_{\mu}\) and, \(D_{\tau}\) respectively, agree with the results reported in \cite{DeSalas2016}.  The results for the coherent function (Eq.~\ref{oscprob}) are shown in figs.~\ref{SLTFBWORe},\ref{SLTFBWORmu} and \ref{SLTFBWORtau}, again for \(D_{e}\), \(D_{\mu}\) and, \(D_{\tau}\) respectively.  Comparing with the results from using the decoherent function, one can find that the results from the coherent function have two distinguishing features.  In the high potential region (\(V_{11} > 10^{-18}~\mathrm{eV}\)), the coherent results for \(D_{\beta}\) have a similar structure to decoherent results, albeit with different values for \(D_{\beta}\) as the coherent results are still baseline dependent, even at high potentials.  As stated before, the decoherent result is the mean accross all baselines of the coherent result.

At lower potentials where the decoherent function produces very small values of \(D_{\beta}\), the coherent function yielded highly variable values.  This suggests that, even for potentials as low as \(10^{-25}~\mathrm{eV}\), a neutrino-Dark Matter interaction could produce a significant shift in the neutrino oscillation spectrum for coherent wavepackets.  This is an improvement in interaction sensitivity over the decoherent result by approximately six orders of magnitude.\footnote{This seemingly coincidental convergence with the amount of improvement in energy resolution required in order to make this measurement is not particularly surprising given that both quantities scale with baseline.  Thus, a one order of magnitude increase in baseline will require a one order of magnitude improvement in both energy resolution and distance measurement, separately, with the benefit of a one order of magnitude improvement in interaction sensitivity.}

\section{Mass Potential}

The mass state interaction Hamiltonian used was:
\begin{equation}
\begin{matrix*}[l]
\mathcal{H}_{tot}&=\mathcal{H}_{vac}+\mathcal{H}_{int}\\
&=
\begin{bmatrix}
E_1 & 0 & 0 \\
0 &  E_2 & 0 \\
0 & 0 &  E_3 \\
\end{bmatrix}+
\begin{bmatrix}
V_{11} & 0 & 0 \\
0 & V_{22} & 0 \\
0 & 0 & 0 \\
\end{bmatrix}
\end{matrix*}
\end{equation}
which, naturally, is the same as used for flavour state interactions save for the exclusion of the PMNS matrix, \(U\).\footnote{Both Hamiltonians are presented in the mass basis representation, which is the canonical basis for the propagation Hamiltonian.}  The parameters used were the same as in the flavour interaction case in order to enable cross comparisons.  The calculations were identical to those for the flavour case.  Due to the nature of the mass state interaction Hamiltonian, the coefficients in the equation for \(f_{\beta}\) are unchanged from the vacuum, which means that the decoherent formula will remain unchanged (i.e. will alays produce a null result).  However, the terms inside the cosines will be changed, meaning that there will be coherence effects.
\begin{figure}
\includegraphics[width=\textwidth,trim={0 3.5cm 0 4.5cm},clip]{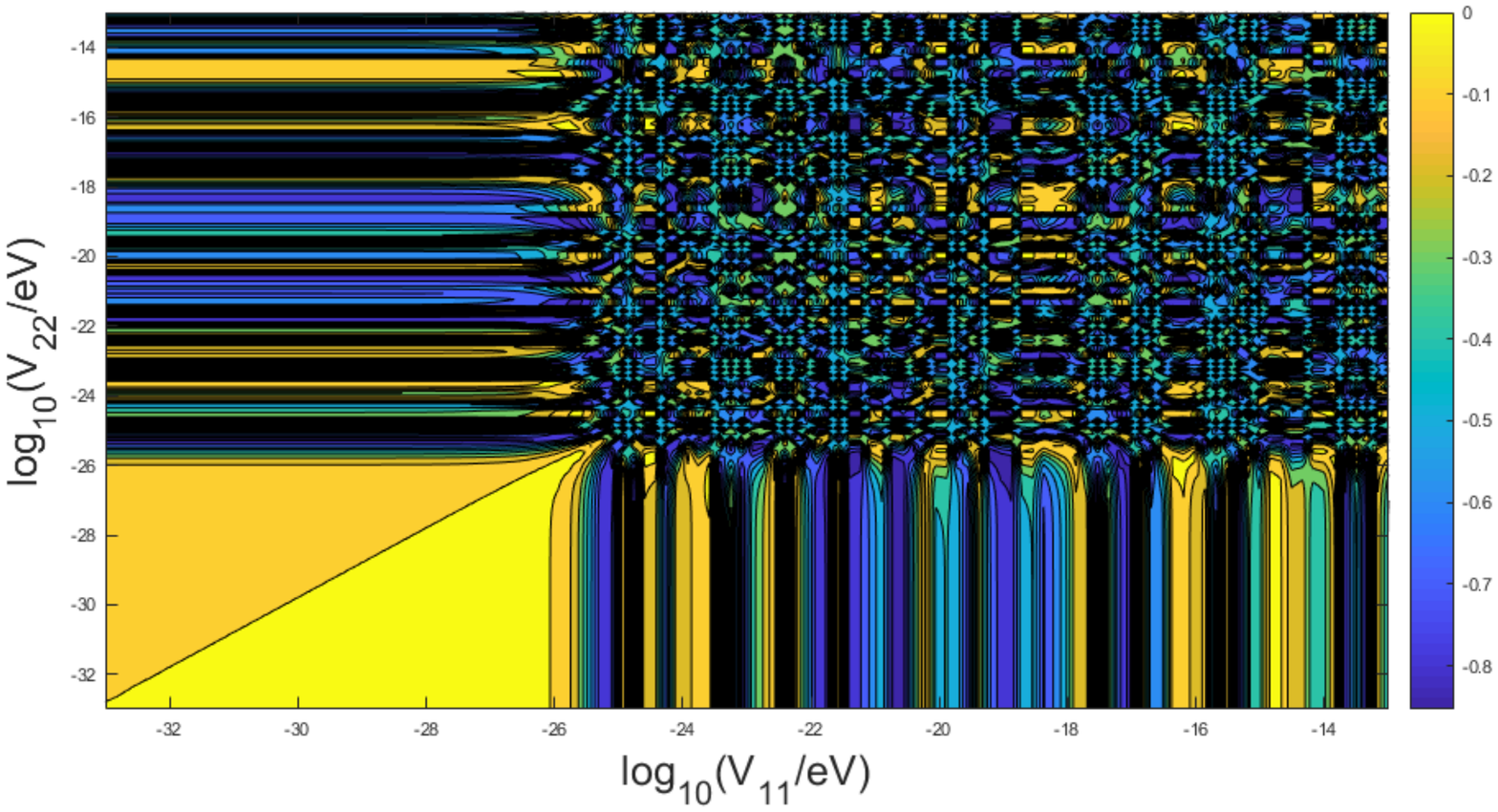}
\caption{This shows \(D_{e}\) across mass potentials at \(1~\mathrm{PeV}\) using the coherent formula.  The baseline was \(L=3{\times}10^{19}~\mathrm{m} \approx 1~\mathrm{kpc}\).}
\label{SLTMBRe}
\end{figure}
\begin{figure}
\includegraphics[width=\textwidth,trim={0 3.5cm 0 4.5cm},clip]{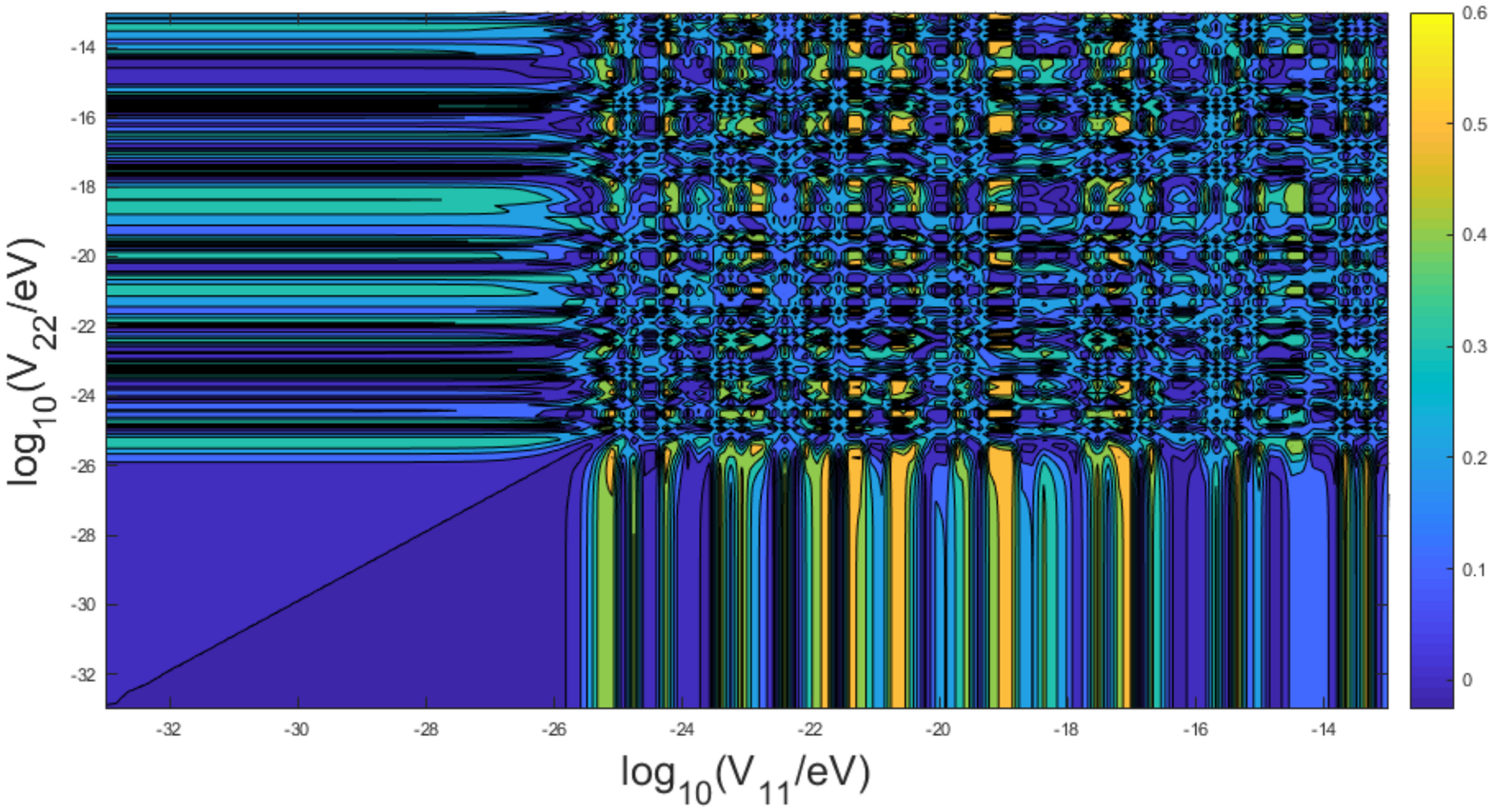}
\caption{This shows \(D_{\mu}\) across mass potentials at \(1~\mathrm{PeV}\) using the coherent formula.  The baseline was \(L=3{\times}10^{19}~\mathrm{m} \approx 1~\mathrm{kpc}\).}
\label{SLTMBRmu}
\end{figure}
\begin{figure}
\includegraphics[width=\textwidth,trim={0 3.5cm 0 4.5cm},clip]{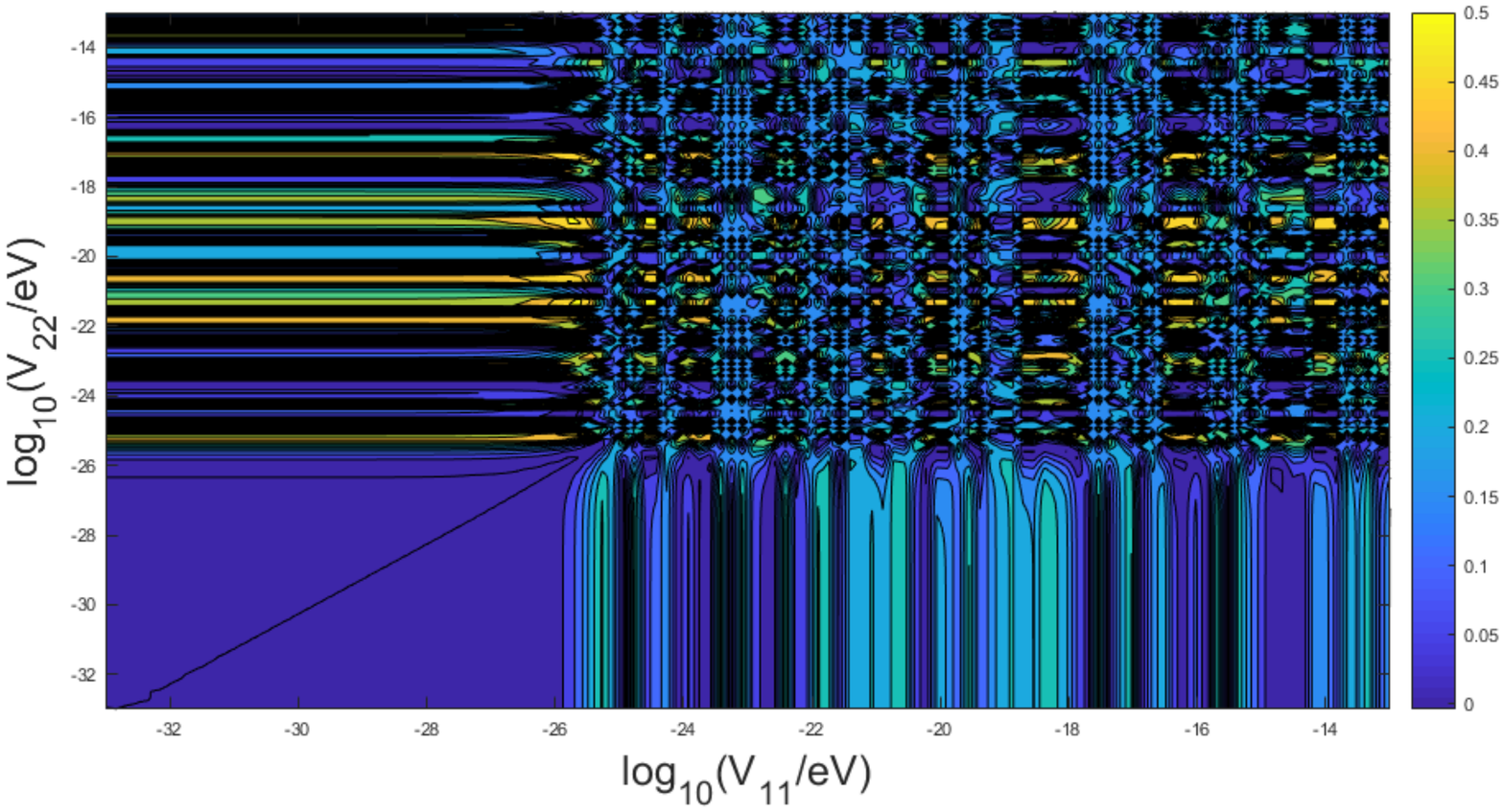}
\caption{This shows \(D_{\tau}\) across mass potentials at \(1~\mathrm{PeV}\) using the coherent formula.  The baseline was \(L=3{\times}10^{19}~\mathrm{m} \approx 1~\mathrm{kpc}\).}
\label{SLTMBRtau}
\end{figure}

The results are shown in figs.~\ref{SLTMBRe}-\ref{SLTMBRtau}.  The patterns produced for the region of \(10^{-26}~\mathrm{eV}<V_{11}<10^{-19}\mathrm{eV}\) are similar to the patterns shown for the same region in the flavour case, including the occurence of significant shifts at potentials lower than can be seen with the decoherent formula.  The mass state results also show more variation in \(D_{\beta}\) for the region \(V_{11}>10^{-19}~\mathrm{eV}\) than the flavour state results.  This shows that the mass state effects consist entirely of the variable small scale coherence effects, as expected.  Another difference between the two cases is that the massive results appear to be entirely symmetric about \(V_{11}\) and \(V_{22}\).  Unfortunately, the minimum cutoff in this spectrum exceeds the limits specified by Lyman-\(\alpha\) forest measurements for fixed cross-sections, which is \(\approx 10^{-38}\) eV\cite{Wilkinson2014}.\footnote{In \cite{Wilkinson2014}, the cosmological constraints are given as cross-sections.  However, a close examination will reveal that the mass of the speculative dark matter particle is included as a parameter.  The potentials are thus a product of the mass-dependent cross-section and the average energy density of DM in the appropriate environment (the galaxy for astronomical neutrinos, the universe for blazar neutrinos).}

\section{Distance dependency}

In neutrino oscillations, there are two different length scales involved which are related to the two mass-squared differences.
The ratio of the two is generally fixed while the total scale is determined by the energy.  In the case of \(1~\mathrm{PeV}\) neutrinos the small scale oscillations are \({\approx}7~\mathrm{AU}\) while the large-scale oscillations are \({\approx}220~\mathrm{AU}\).
\begin{figure}
\includegraphics[width=\textwidth,trim={0 3.5cm 0 4.5cm},clip]{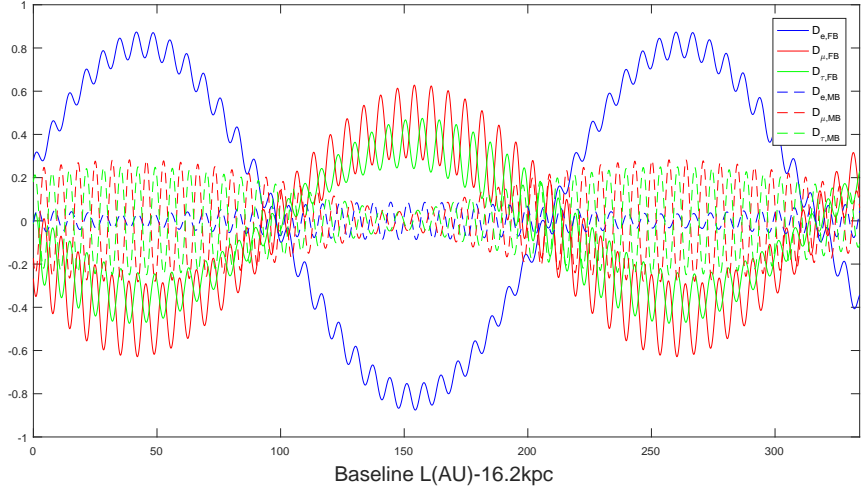}
\caption{This shows the difference patterns for all three flavours in both potentials after a displacement of 16.2 kpc from the source with the baseline in AU.  The two potentials (\(V_{11}\) and \(V_{22}\)) were both set to \(10^{-26}~\mathrm{eV}\).}
\label{SLTBSLNSRO}
\end{figure}
The presence of an interaction introduces a shift in the oscillations.  The shift is, first and foremost, in frequency, which causes the oscillations in the presence of an interaction to be out of phase with the oscillations in the absence of the interaction.  The shift can also be more subtle, which seems to occur with the mass state interaction as the maximum differences are not as large as those for flavour (see fig.~\ref{SLTBSLNSRO}).

\begin{figure}
\includegraphics[width=\textwidth,trim={0 3.5cm 0 4.5cm},clip]{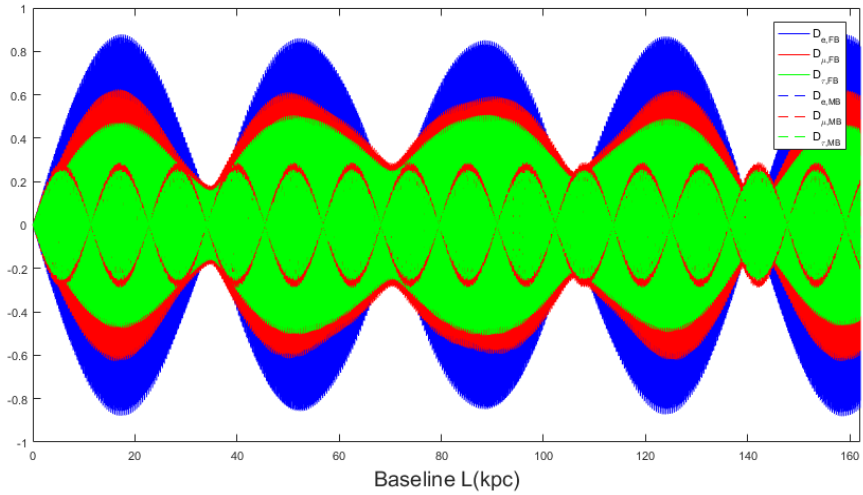}
\caption{This shows the long range difference patterns for all three flavours with both potentials.  The small amplitude oscillations are for the mass case and the large amplitude oscillations are for the flavour case.  The potentials are the same as before.  This result is of strictly academic interest as the potentials are considered unphysical at these distances.}
\label{SLTBSLNLRO}
\end{figure}

An interesting difference between the two interactions is the effect on the different flavours (see fig.~\ref{SLTBSLNLRO}).  In the flavour case the biggest effect is on the electron neutrinos while in the mass case the effect on electron neutrinos is quite small.  Another interesting effect is how the long range pattern for the flavour case shows a slight shift in the shape of each peak, which shows up most clearly for muon neutrinos.  Due to the unphysical nature of the combination of baseline and potential (potential is related to dark matter concentration) any discussion of the long range effects are purely academic, yet this is still an interesting result.

\section{Results for Blazar Neutrinos}
A very exciting recent developement in neutrino physics was the first identification of a point source of astrophysical neutrinos.  The IceCube Collaboration identified blazar TXS 0506+056 as the source for a number of high-energy (290 TeV) neutrinos observed by the group over a period of 9.5 years.  In this section we present the results for \(D_{\beta}\) in the scenario where only electron neutrinos are produced at TXS 0506+056.

Blazar TXS 0506+056 has a redshift of \(z=0.3365 \pm 0.0010\), which corresponds to a baseline of \((4.384 \pm 0.013)\times 10^{25}~\mathrm{m}\) \cite{2041-8205-854-2-L32}). The energy was set to be 290 TeV as that was the energy reported by \cite{Collaboration2018}.  The potential range was changed to be in the range \(10^{-40}~\mathrm{eV} \sim 10^{-13}~\mathrm{eV}\) in order to accommodate the new results.
\begin{figure}
\includegraphics[width=\textwidth,trim={0 3.5cm 0 4.5cm},clip]{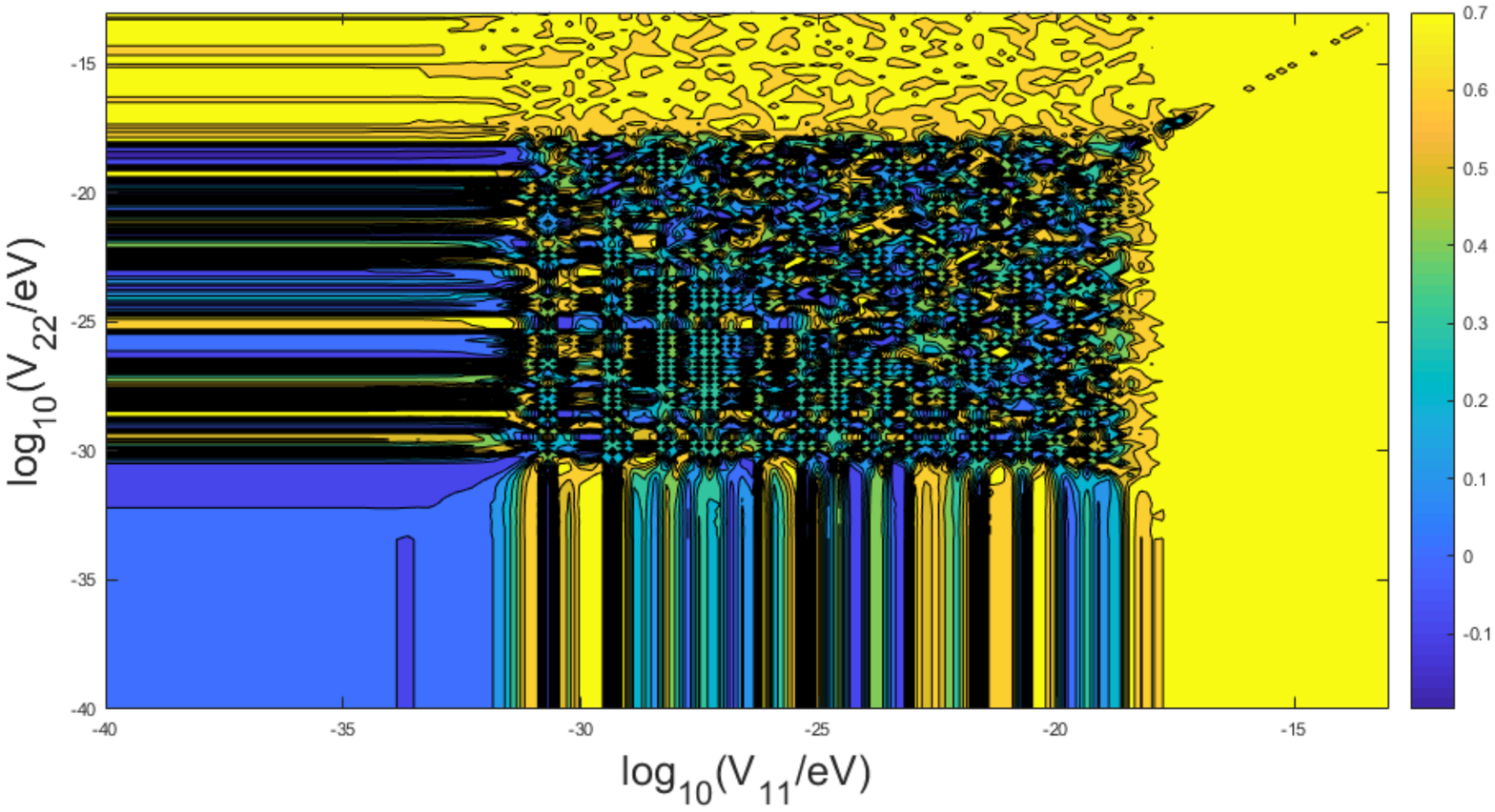}
\caption{This shows \(D_{e}\) in the flavour basis at \(290~\mathrm{TeV}\) using the coherent formula.  The baseline was \(L=4.384{\times}10^{25}~\mathrm{m}\).  The two kinds of effects are both still visible.  The biggest difference from previously is that the region of high variability extends farther.  This is due to the longer baseline.}
\label{SLTFBWORenew}
\end{figure}

The main effect was from the baseline increase, which dramatically extends the range of observability.  The observable cutoff have now been lowered to  \(V_{11} \approx V_{22}  \approx 10^{-33}\) eV.
This is understandable as a longer baseline means that there is more Dark Matter for the neutrinos to scatter off of.  However, at this baseline, detection of coherent neutrino states would be extremely difficult, requiring a much smaller energy resolution than the scenarios discussed in sections 3 and 4.  Essentially, these results correspond to an ideal detector detecting neutrinos emitted by an ideal blazar and are thus little more than a thought experiment.

\section{Discussion}

The work by \cite{DeSalas2016} showed possible ``Dark Matter effects" on the astronomical neutrino spectrum.  This work showed that coherence effects can dramatically increase the sensitivity of the spectrum to dark matter interactions.


The first major issue is that of coherence, specifically whether coherent astrophysical neutrino states can be detected anytime in the forseeable future.  This would require neutrinos produced at not too great a distance at rather high energies detected by more sophisticated equipment than currently exists.  Additionally, the object in question would require a physical size substantially less that the oscillation wavelength.  As previously stated, these constraints imply that the most likely source is in the near vicinity of stellar-mass black holes.  It is an open question as to whether stellar-mass black holes are indeed potential sources of sufficiently high energy neutrinos.  How such neutrinos could be linked to a specific source is also an issue that needs to be addressed.  Then there is the constraint on baseline uncertainty.  Given that this is fixed by the neutrino energy and that black hole range-finding is somewhat difficult, overcoming this obstacle is arguably even more difficult than overcoming the energy resolution constraint.  On the other hand, this question is of interest to other fields so there might be faster progress in this area.

Even if the source was known and the coherence constraints could be overcome, there is the issue of detecting enough neutrinos to be able to conclude anything about the flavour spectrum, an issue which is compounded by the large distances involved and the inverse square law.  The use of very high energy neutrinos helps somewhat, as these neutrinos are not produced in the Solar System, which reduces the background that needs to be accounted for (relative to lower energy neutrinos).  Still, there needs to be a sufficiently large signal.  For blazar neutrinos, there has, as of writing, only been a handful of confirmed events \cite{Collaboration2018}.  This, coupled with the issues regarding detector energy resolution, means that this type of analysis cannot be performed in the near future (which most emphatically does not mean that it cannot be performed in the distant future).

Unfortunately, the ranges for both interactions exceed the cosmological bound for a constant cross-section.  This disfavours the mass state interactions as those interactions are predicted to have constant cross-sections.  The flavour state interactions are predicted to be energy dependent like the SM weak interaction, and energy dependent interactions are far less bound by cosmological constraints\cite{Wilkinson2014}, meaning that a flavour state interaction is, perhaps unsurprisingly, more likely to be detected by this kind of experiment (presuming that one even exists and is sufficiently strong).  

If such an effect was observed it would yield valuable physics information.  Depending on the type of interaction and the potential at which it occured (a distinction that can't be made with the de-coherent effect), it woud help to fix the absolute mass scale of the DM particle.  Also, such experiments would complement cosmological observations rather well given the different energy scales being observed.

All of that being said, given both the need for a large sample size and significant improvements in detector quality, it will probably be a very long time before making these kinds of observations becomes feasible.

\section{Summary}

Calculations for astronomical neutrino spectra were performed using both the coherent and decoherent formulas for neutrino flavour transition probabilities.  This was done for both the flavour and mass state interactions.  The results show that using the coherent formula can produce larger effects at lower potentials in the flavour case and is essential for obtaining any effect in the mass case.  An examination and comparison of the baseline dependency of the coherent formula in each basis showed that it should be possible to distinguish between the two bases, which would shed light on the nature of both neutrinos and Dark Matter.  Similar calculations were performed with parameters related to high energy neutrinos produced by blazar TXS 0506+056 which showed that the observable effect increases with baseline.  The overall result is that coherence effects dramatically improve the detectability of neutrino-DM interactions.  However, the detection of coherent astronomical neutrinos presents a significant challenge to experimentalists since such a detection would require an improvement in energy resolution by approximately six orders of magnitude (or more) in both energy resolution and astronomical distance measurement precision, which might be hard to achieve in the near future.


\end{document}